\documentclass[letterpaper,twoside,english,reprint]{revtex4-2}
\usepackage{lmodern}

\usepackage[T1]{fontenc}
\usepackage[latin9]{inputenc}
\usepackage{color}
\usepackage{babel}
\usepackage{amsmath}
\usepackage{amssymb}
\usepackage{graphicx}
\usepackage[unicode=true,pdfusetitle,
 bookmarks=true,bookmarksnumbered=false,bookmarksopen=false,
 breaklinks=false,pdfborder={0 0 1},backref=false,colorlinks=true]
 {hyperref}
\hypersetup{
 linkcolor=blue, urlcolor=blue, citecolor=blue}

\makeatletter

\pdfpageheight\paperheight
\pdfpagewidth\paperwidth

\usepackage{hyperref}
\usepackage[all]{hypcap}

\makeatother

\begin{document}
\title{When can few-species models describe dynamics within a complex community?}
\author{Stav Marcus and Guy Bunin}
\affiliation{Department of Physics, Technion - Israel Institute of Technology,
Haifa 32000, Israel}
\begin{abstract}
Dynamics of species' abundances in ecological communities are often
described using models that only account for a few species. It is
not clear when and why this would be possible, as most species form
part of diverse ecological communities, with many species that are
not included in these few-variable descriptions. We study theoretically
the circumstances under which the use of such models is justified,
by considering the dynamics of a small set of focal species embedded
within a diverse, sparsely-interacting community. We find that in
some cases the focal species' dynamics are high-dimensional, making
a few-variable description impossible. In other cases we show that
such a description exists, even though the effect of the surrounding
community on the focal species' dynamics is not small or simple. We
give two different methods for approximating the dynamics, by using
effective parameters that depend on the surrounding community, which
are relevant under different assumptions on the relation between the
explicitly modeled focal species and the rest of the species. Both
methods work surprisingly well in many of the cases that we check,
with effective dynamics that are often very similar and sometimes
indistinguishable from the true dynamics, even when the effect of
the community on the focal species is significant.
\end{abstract}
\maketitle

\section{\label{sec:Introduction local description}Introduction}

In community ecology, theoretical models that consider only a few
species are commonplace, often provide theoretical insight, and have
successfully accounted for a range of natural phenomena. Classic textbook
examples include phenomena such as competitive exclusion between competing
species and predator-prey cycles \citep{may_theoretical_2007}. The
success of such models in natural situations may seem surprising when
describing a handful of species that form only a part of a diverse
community harboring many interacting species, and it is therefore
natural to ask when and why few-species descriptions are applicable
to such situations.

Here we consider this issue in the context of persistent fluctuations
in species populations, a ubiquitous and important phenomenon \citep{lundberg_population_2000,inchausti_relation_2003}.
Persistent fluctuations have been observed in few-species communities
in experiments and in the field \citep{elton_periodic_1924,utida_cyclic_1957,sinervo_rockpaperscissors_1996,peterson_wolfmoose_1999,bonsall_demographic_2004},
and have been analyzed using few-species models predicting, for example,
population cycles \citep{lack_natural_1954,may_limit_1972,beninca_species_2015}.
This is despite the fact that in some cases, the dynamics have been
shown to be higher dimensional \citep{stenseth_population_1997,begon_predatorprey_1996}.

\begin{figure*}
\begin{centering}
\includegraphics[width=1\textwidth]{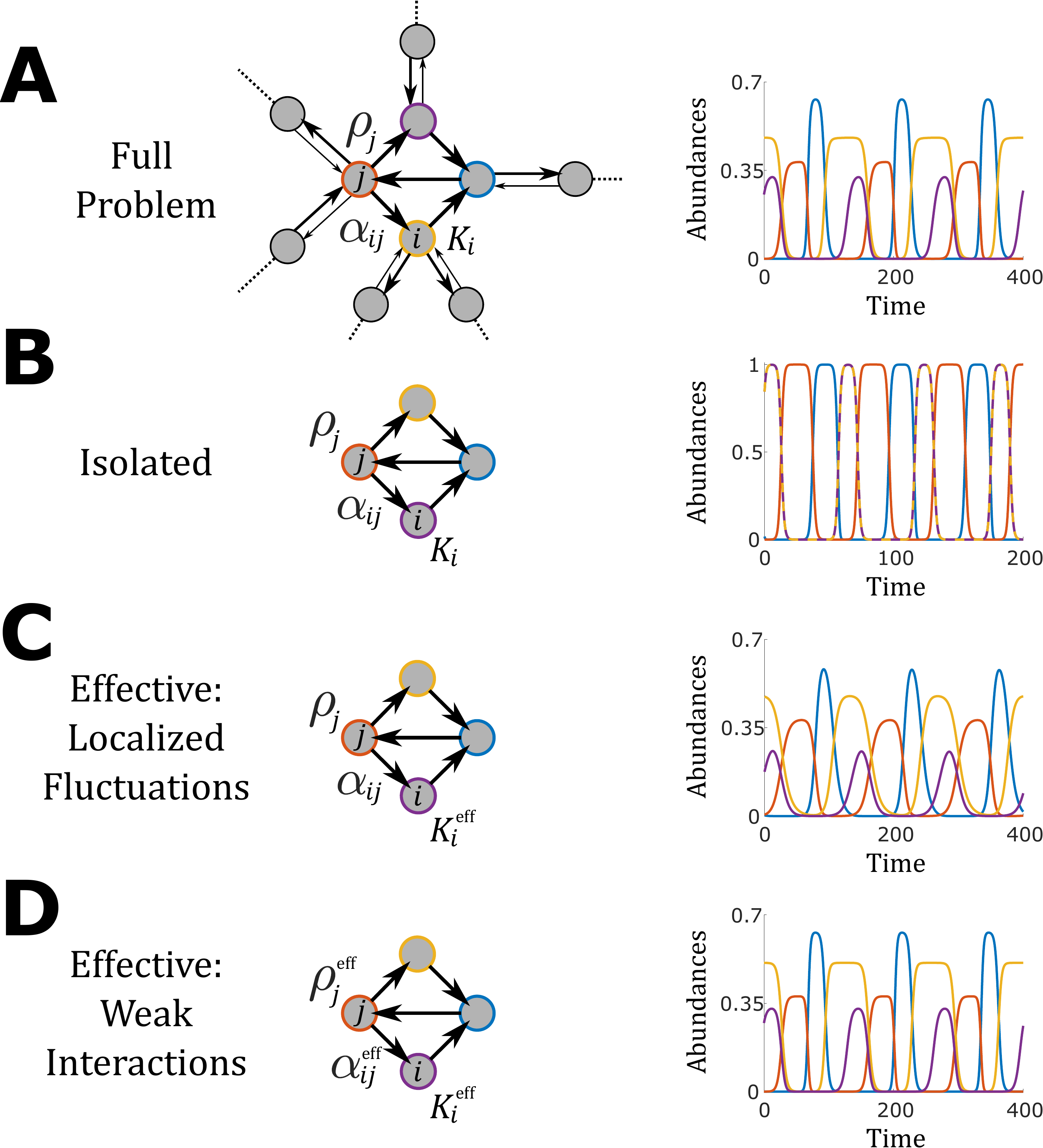}
\par\end{centering}
\caption{\label{fig:setup}\textbf{An example for the general setup for approximating
the behavior of a few species within a diverse community. }On the
left are the interactions between the focal species. Thicker edges
denote stronger interactions. On the right are the abundance dynamics
$N_{i}(t)$ of the focal species, at long enough times by which the
dynamics have stabilized to a limit cycle. The color of each vertex
matches the corresponding abundance plot. \textbf{(A)} The full problem:
a handful of focal species form part of a large ecological community,
interacting with other species in it. The interactions can be bidirectional
and asymmetric. The dynamics between the focal species are parametrized
by the interaction strengths $\alpha_{ij}$, the carrying capacities
$K_{i}$ and the growth rates $\rho_{j}$. \textbf{(B-D) }Different
few-variable models yielding effective dynamics for the focal species.
\textbf{(B)} The species are considered in isolation, using the same
parameters for the dynamics in the community. Typically, this gives
dynamics that are very different from the ones observed in the community.\textbf{
(C-D) }Only the focal species are modeled, but using different \textquotedblleft effective\textquotedblright{}
parameters. We suggest two possible sets of effective parameters,
which can yield different results. \textbf{(C)} Using the decaying
fluctuations approximation, one finds effective $K_{i}^{\text{eff}}$
while using the same values for $\alpha_{ij}$ and $\rho_{j}$. \textbf{(D)}
Using the weak interactions approximation, one find effective values
for all parameters. Both (C,D) result in a much better approximation
of the embedded dynamics than using the same parameters as in the
full community (B).}
\end{figure*}

In this paper, we will discuss circumstances under which few-species
models can approximate the true dynamics of the species in nature.
We consider a subset of ``focal species'' that interacts with other
species in the community, see Fig. \ref{fig:setup}(A). Collectively,
we will refer to these species as the ``focal group''. We consider
cases where in isolation, the abundances of the focal species vary
in time, see Fig. \ref{fig:setup}(B). One can then conceive of three
possibilities:

(1) The dynamics of the focal species are essentially unaffected by
the rest of the community, and so they behave as if the other species
are not present at all. Model parameters can then be found that are
independent from the rest of the community. Below we find that this
scenario is often too simplistic, and even rather weak interactions
with the rest of the community can significantly alter the dynamics,
see Fig. \ref{fig:setup}(A,B).

(2) The dynamics of the focal species are significantly affected by
the rest of the community, however \emph{there does exist} a few-variable
set of equations that captures the dynamics to a good approximation,
see Fig. \ref{fig:setup}(C,D). The model parameters describing the
dynamics in the few variable description (interaction coefficients,
carrying capacities, etc.) are not the ``true'' parameter values
that would be used for a dynamical model that explicitly follows all
species the in the entire community. Rather, these are effective parameters,
that account indirectly for the interactions with the other species.
This possibility is most relevant and applicable when the fluctuations
are concentrated on the focal species and some species surrounding
them.

(3) The dynamics cannot be captured by a few-species model. This can
happen when many species in the community fluctuate, and affect the
focal species in complex ways. We studied the conditions for fluctuations
of a few species, versus fluctuations of many of them in another work
\citep{marcus_local_2023}, and briefly review the results below.

\begin{figure*}[p]
\begin{centering}
\includegraphics[width=1\textwidth]{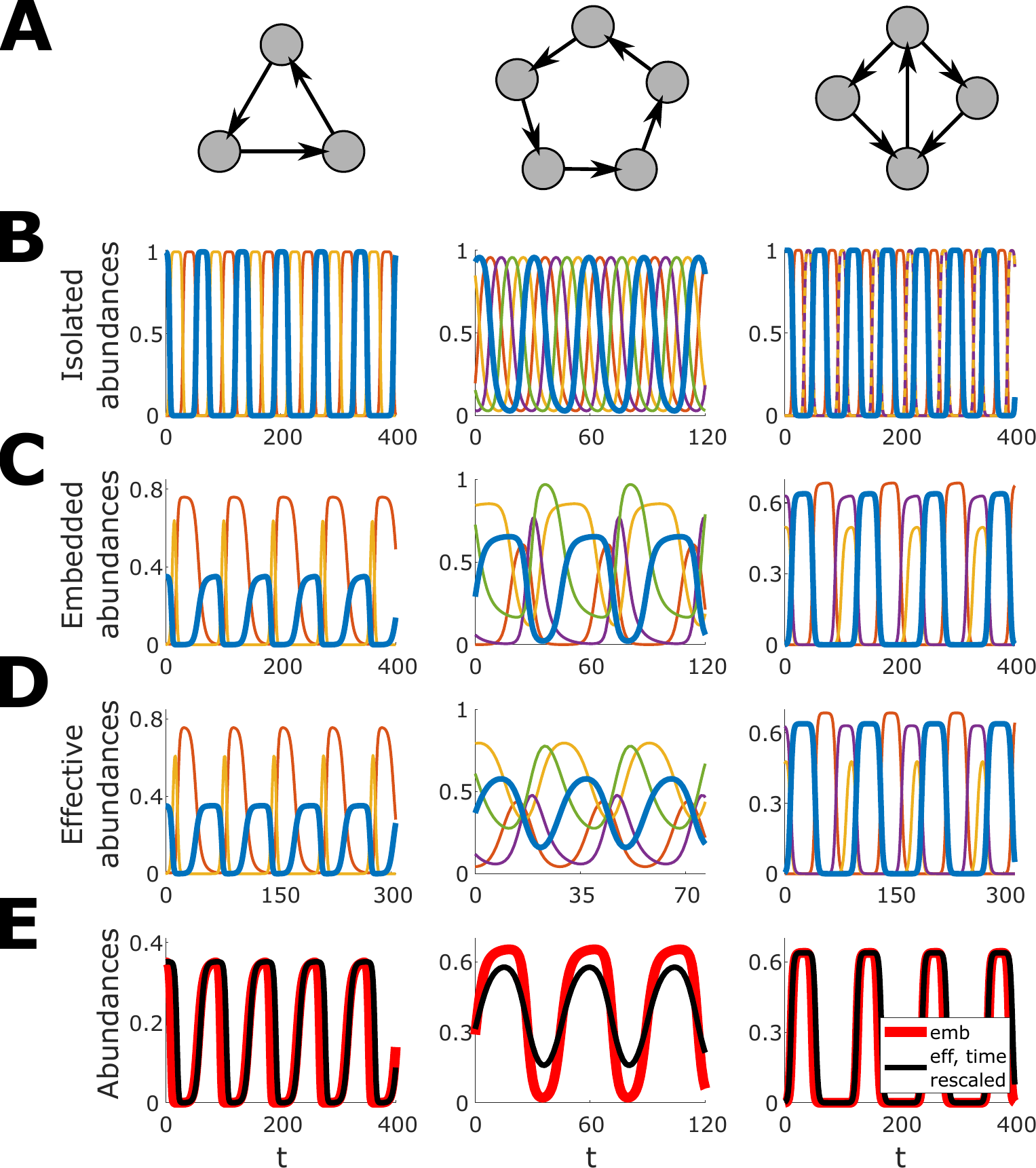}
\par\end{centering}
\caption{\label{fig:examples}\textbf{Comparing embedded and effective dynamics.
}Examples of the dynamics of three different focal groups, where each
column gives a different example. \textbf{(A)} The focal group. \textbf{(B)}
The dynamics of all focal species when in isolation. \textbf{(C)}
The dynamics of the focal species when embedded in the community.
\textbf{(D)} The effective dynamics of the focal species found using
the Localized fluctuations approximation at distance $D=0$. The dynamics
here are shown with a different timescale, as without time-rescaling
the effective dynamics have a different frequency than the embedded
dynamics. \textbf{(E)} Comparison of the dynamics of one of the focal
species, when embedded in the community (thick red line) and in the
effective description with time rescaling (black). The abundance curve
of the species shown here is shown as a thick blue line in all rows
2-4. Details on parameters are given in Appendix \ref{sec:Simulation-details}.}
\end{figure*}

The focus in this paper will be on possibility number 2 above, asking
when a model that only explicitly accounts for the focal species can
capture the dynamics of the focal species within the entire community,
and how to find the parameter values for such a model. We consider
a community that harbors many species, and assume that the entire
community follows Lotka-Volterra dynamics, and that the interaction
network is predominantly competitive and sparse, so that each species
interacts significantly with only a handful of the many others. We
suggest two methods for constructing few-species models and assigning
effective parameters in these models. The two approaches work under
different assumptions, and both the number of species explicitly modeled
and and the effective parameters assigned to them can differ between
them. The first method is an expansion around weak feedback from the
species surrounding the focal species as a result of weak interactions.
The second is an expansion in the size of the temporal fluctuations,
assuming that they are small for the surrounding species, and localized
in the interaction graph around the focal group. We will show that
in many cases, the effective dynamics obtained using few-species dynamics
with these approximations are extremely similar to the dynamics of
the focal species in the full community, see examples in Fig. \ref{fig:examples}.
The general setup of our work is shown in Fig. \ref{fig:setup}.

\subsection{The model}

We model the community dynamics using the generalized Lotka-Volterra
(LV) equations, where the dynamics of the abundance $N_{i}$ of species
$i$ is governed by:

\begin{equation}
\dot{N}_{i}=\rho_{i}N_{i}\left(K_{i}-\sum_{j}\alpha_{ij}N_{j}\right)+\lambda\equiv N_{i}g_{i}(\vec{N})+\lambda\,.\label{eq:LV}
\end{equation}
$\alpha_{ij}$ are the interaction strengths, with the intraspecific
interactions $\alpha_{ii}=1$. $K_{i}$ are the carrying capacities,
and the bare growth rates are $r_{i}=\rho_{i}K_{i}$. We take $K_{i}=1,\rho_{i}=1$
in all simulations. $\lambda$ is a migration rate from a regional
species pool with $S$ species, assumed to be much smaller than any
other scale in the system, and for simplicity taken to be identical
for all species, which has little effect on the main results. Due
to the migration, the abundance of an extinct species, i.e. with $N_{i}\simeq0$
(up to small values of order $\lambda$), can grow again if it has
a positive growth rate, $g_{i}(\vec{N})>0$.

We consider a set of a focal species (between 3 and 5 in the examples
below), embedded in a larger community. We test different interaction
networks between the focal species, which in isolation fluctuate indefinitely
in a limit cycle. The focal species are embedded in a community of
$S$ non-focal species, typically many more than the number of focal
species. The interactions in the community are assumed to be sparse,
sampled at random such that the number of interactions for each species
is Poisson distributed with mean $C$ ($C$ is known as the average
degree, or connectedness, of the network).

In simulations, we use interactions of constant strength: for the
non-zero links, we take $\alpha_{ij}=\alpha$ in one direction, and
$\alpha_{ji}=\beta$ in the other direction. We also consider the
unidirectional case, where $\beta=0$. The embedding of the focal
species is done by taking each interaction (incoming or outgoing)
between a focal species and a non-focal community species to be $\alpha$
with probability $C/2S$, and zero otherwise. After all nonzero interactions
are drawn, we take their reciprocal interactions to be $\beta$. The
interactions between the non-focal species in the embedding community
are sampled in the same manner. We also take all $K_{i}=1,\,r_{i}=1$.
This is the model that we use in \citep{marcus_local_2023}, except
for a slightly different definition for $C$ (as we took interactions
as nonzero with probability $C/S$).

In the following, the dynamics of the focal species found by solving
the full Lotka-Volterra equations (Eq. \ref{eq:LV}) for all species
(focal and non-focal) will be referred to as the ``embedded dynamics''.
These are the true dynamics of the species in nature, where they form
a part of the entire community. The dynamics found using few-species
models with effective parameters will be referred to as the ``effective
dynamics''.

In \citep{marcus_local_2023}, we studied the dynamics of communities
with random sparse interaction networks. We found that they exhibit
four qualitative dynamical behaviors, termed ``dynamical phases'',
depending on the connectivity $C$ and the symmetry of the interactions
$\alpha_{ij}$ (namely, whether $\alpha_{ij}$ and $\alpha_{ji}$
have similar values). For interactions that are far from symmetric,
for example unidirectional interactions (such that $\alpha_{ij}\neq0\Rightarrow\alpha_{ji}=0$),
we find three phases. The first is a fixed-point phase, where the
abundances of all species always relax to a fixed point after long
enough times. The second, a local-fluctuations phase, where some species
abundances fluctuate indefinitely. These fluctuations originate in
a local structure in the interaction graph, and involve only a finite
number of species even as the system size $S$ becomes infinite. Finally,
there is a an extensive fluctuations phase, where a finite \emph{fraction}
of species abundances undergo persistent chaotic dynamics. Importantly,
these dynamics are high-dimensional and cannot be stopped by the removal
of any finite number of species from the community. This means that\emph{
there can be no local description of the dynamics} \emph{in the extensive
fluctuations phase}, corresponding to possibility number 3 in Section
\ref{sec:Introduction local description}. We will therefore not consider
it any further in the present paper. When interactions are symmetric
($\alpha_{ij}=\alpha_{ji}$) or close to that, the dynamics again
always reach a fixed point \citep{pykh_lyapunov_2001}.

\section{Results}

\subsection{Unidirectional interactions: an exact local description}

\begin{figure*}
\begin{centering}
\includegraphics[width=1\textwidth]{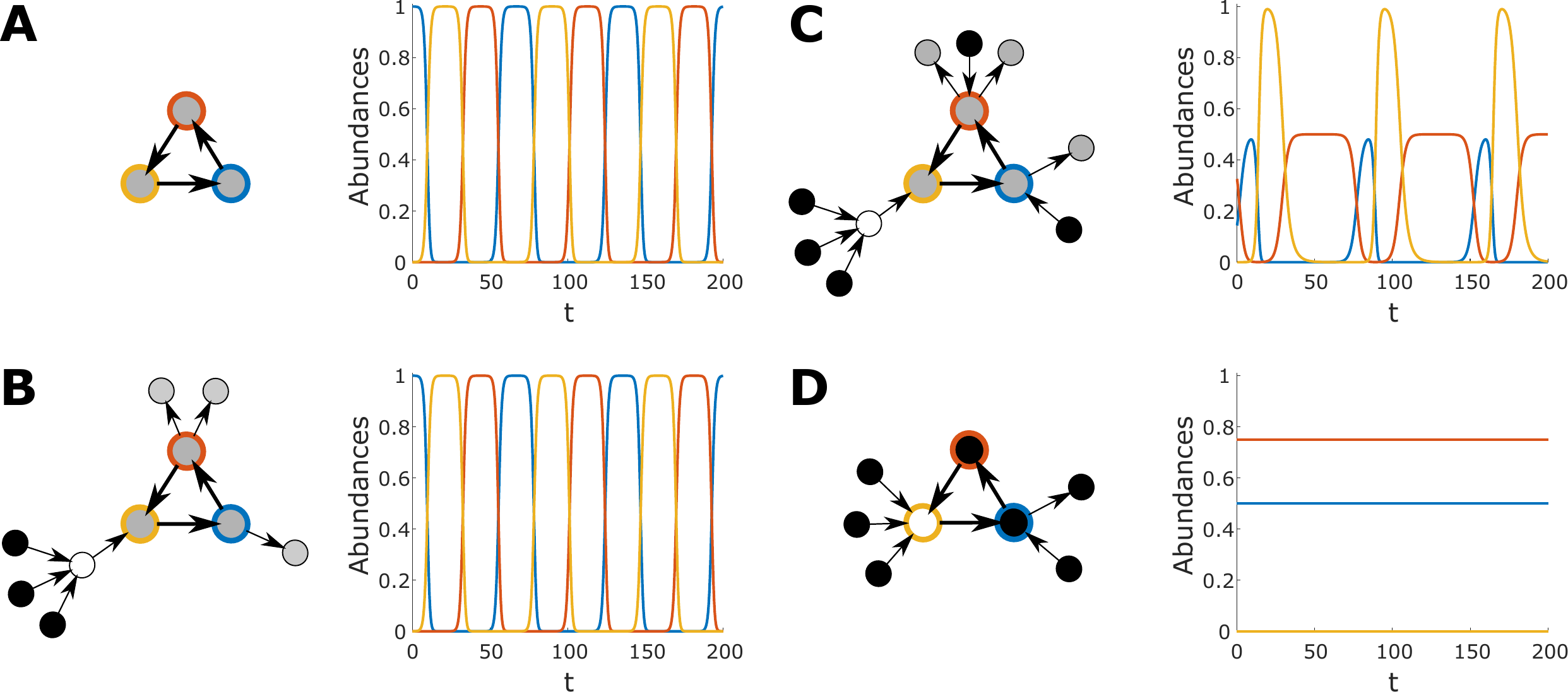}
\par\end{centering}
\caption{\label{fig:unidirectional example}\textbf{Dynamics of focal species
embedded in a network with unidirectional interactions.} Simple examples
where the embedding community includes only a few species, in order
to illustrate the mechanisms at work. The focal group is a unidirectional
cycle of length 3 and with interaction strength $\alpha_{\text{focal}}=2.5$,
and the (small) embedding community has $\alpha_{\text{community}}=0.5$.
The dynamics of the focal species are shown next to matching directed
graphs representing the \emph{entire system.} In the graph, extinct
species ($N_{i}=0$) are shown in white, species with a fixed abundance
$N_{i}>0$ in black, and species with fluctuating abundances in gray.
The color of the border of a vertex representing each of the focal
species matches the corresponding abundance plot. \textbf{(A) }The
abundances of the focal species fluctuate in isolation. \textbf{(B)
}The rest of the community does not affect the focal species. The
only interactions between focal group and the community are either
outgoing from the focal group or incoming from extinct species. The
dynamics are identical to the isolated case. \textbf{(C) }Two of the
focal species (blue and red) are affected by the rest of the community
through an incoming interaction from species with $N_{j}=1$, reducing
their effective carrying capacities. The abundances fluctuate with
very different dynamics from the isolated dynamics. \textbf{(D) }One
of the focal species (yellow) is driven extinct by the rest of the
community, as it has three incoming interactions from species with
$N_{j}=1$. The abundance dynamics reach a fixed point.}
\end{figure*}

We will begin the discussion in the case where interactions in the
embedding community are unidirectional: for any $\alpha_{ij}>0$,
the reciprocal interaction is $\alpha_{ji}=0$. Considering cases
where the focal species in isolation exhibit persistent fluctuations,
their dynamics when embedded in the community might be the same as
when isolated, fluctuate indefinitely with different dynamics, or
reach a fixed point. See examples in Fig. \ref{fig:unidirectional example}.

In all these cases, we construct a dynamical model describing only
the focal species, which exactly captures the embedded dynamics when
the surrounding community contains many species ($S\gg1$). In this
many-species limit, the abundances of species with incoming interactions
to the focal group are fixed and uncorrelated with each other. This
is due to the following:
\begin{enumerate}
\item \emph{In sparse systems, short loops are rare and the neighborhood
of any species is tree-like with probability one \citep{barkai_properties_1990}.}
\item \emph{Element fluctuations have no feedback effects}: The spread of
fluctuations from the focal species in the parameter regimes we consider
is finite by definition. In the local fluctuations phase the downstream
effect of fluctuations is reaches only a finite number of species.
In the fixed point phase, the amplitude of fluctuations is reduced
by a factor of $\alpha$ in each step of the downstream spread \citep{marcus_local_2023}.
Therefore, feedback effects would require a short directed loop beginning
and ending in the element, which from the previous point has probability
zero.
\item \emph{Before the embedding, species interacting with the focal species
have fixed abundances}: In the phases considered the system is either
at a fixed point or has at most a finite number of fluctuating species,
and therefore the probability that the element will be randomly connected
to a fluctuating species is zero in the limit $S\rightarrow\infty$.
\end{enumerate}
The incoming interactions to the focal species effectively reduce
their carrying capacity by a constant value. The dynamics for the
abundance $N_{i}$ of species $i$ of the element can be written as

\begin{align}
\dot{N}_{i} & =\rho_{i}N_{i}\left(K_{i}-\sum_{j\text{\ensuremath{\notin}focal}}\alpha_{ij}N_{j}-\sum_{j\text{\ensuremath{\in}focal}}\alpha_{ij}N_{j}\right)+\lambda\\
 & \equiv\rho_{i}N_{i}\left(K_{i}^{\text{eff}}-\sum_{j\text{\ensuremath{\in}focal}}\alpha_{ij}N_{j}\right)+\lambda\,,\nonumber 
\end{align}
defining an effective carrying capacity
\begin{equation}
K_{i}^{\mathrm{eff}}=K_{i}-\sum_{j\text{\ensuremath{\notin}focal}}\alpha_{ij}N_{j}\,.\label{eq: k_eff constant}
\end{equation}
Importantly, $K_{i}^{\mathrm{eff}}$ is a constant: as explained above,
the abundances of species with incoming interactions to the element
are fixed in the limit $S\rightarrow\infty$. Therefore, taking the
element in isolation and replacing $K_{i}\rightarrow K_{i}^{\text{eff}}$
is \emph{an exact description} of the long term dynamics. Note that
$\rho_{i}=r_{i}/K_{i}$ does not change, which amounts to taking $r_{i}\rightarrow r_{i}^{\text{eff}}=\left(K_{i}^{\text{eff}}/K_{i}\right)r_{i}$.
Focal species that are driven extinct by the embedding community have
a negative $K_{i}^{\mathrm{eff}}$, and go extinct when using the
effective dynamics as well, see Fig. \ref{fig:unidirectional example}(D).
The reduced values of the carrying capacities can be seen in Fig.
\ref{fig:unidirectional example}(C): two of the species have reduced
effective carrying capacities, the maximal abundances they reach is
close to their $K_{\text{eff}}=0.5$; the carrying capacity of the
third species is unaffected by the community, and so it reaches a
maximal abundance close to $K_{i}=1$.

One special case occurs when all nonzero interactions in the embedding
community, and between the community and the focal species, have $(K_{j}/K_{i})\alpha_{ij}>1$
(in the simple model we use in simulations, the condition is $\alpha>1$).
In this case, all non-focal species (possibly except for a finite
number) are either extinct with $N_{i}=0$ or have $N_{i}=K_{i}$
\citep{marcus_local_2023}. Therefore, any focal species $i$ with
an incoming interaction from a non-focal species $k$ with $N_{k}=K_{k}$
will have a negative growth rate: $g_{i}(\vec{N})\leq\rho_{i}(K_{i}-\alpha_{ik}N_{k})<0$.
Therefore, in this case there are two options. In the first, some
of the focal species are driven extinct by the environment, and as
a result its dynamics will be changed, possibly reaching a fixed point
rather than fluctuating. In the second, the focal species are unaffected
by the rest of the community, and has the same dynamics as they would
have in isolation. In both cases, the effective dynamics are given
by considering the focal species as if in isolation, while removing
any of them that are driven extinct by the rest of the community.

\subsection{Bi-directional interactions}

When interactions are bi-directional, there are feedbacks between
the focal species and their neighborhood in the network. Therefore,
if the focal species abundances fluctuate, the abundances of its neighbors
can fluctuate as well, and one can no longer assume that the effect
of incoming interactions will amount to a fixed change in the carrying
capacities $K_{i}$.

Here there are three possible cases:
\begin{enumerate}
\item The interaction with the community dampens the fluctuations, so that
the entire community reaches a fixed point (see Fig. \ref{fig:statistics}A).
In such a case, as all species interacting with the focal species
have fixed $N_{i}$, and one can find an effective description by
taking $K_{i}^{\text{eff}}$ as in Eq. \ref{eq: k_eff constant},
which gives an exact description of focal species' dynamics.
\item There are fluctuations, and the system is in the local fluctuations
phase. The fluctuations will only spread to a finite number of species,
as they will at some point be ``cut off'' by extinct species (otherwise,
if fluctuations spread further the system would be in an extensive
fluctuation phase). While an exact description of the element alone
may not be possible, one can be found by expanding the description
to the focal species and others in the isolated sub-component of the
network.
\item There are fluctuations, and the system is in the fixed point phase.
Here there is no sharp isolation of the embedded structure, but fluctuations
decay exponentially with the distance in the interaction graph from
the fluctuating center. This remaining case will be the focus below.
\end{enumerate}
In the following, we will use two measures for the accuracy of the
approximations. The first is whether both the effective and embedded
dynamics reach a fixed point or continue to fluctuate. Note that if
the embedded dynamics reach a fixed point, this is option 1 above,
for which the effective description is exact, so the effective dynamics
will also reach a fixed point. Therefore, the probability of the effective
dynamics to fluctuate is always smaller or equal to that of the true,
embedded dynamics. The other measure will be the difference $\Delta$
between the embedded abundances $\left\{ N_{i}(t)\right\} $ and the
effective abundances $\left\{ N_{i}^{\text{eff}}(t)\right\} $, which
we define as 
\begin{equation}
\Delta=\frac{1}{\left|\text{focal group}\right|}\sum_{i\in\text{focal group}}\left\langle \left|\left(N_{i}(t)-N_{i}^{\mathrm{eff}}(t)\right)\right|\right\rangle \,,
\end{equation}
where the brackets $\left\langle \cdot\right\rangle $ denote average
over time, at long enough times after the dynamics relax to a limit
cycle or a fixed point.

\begin{figure*}
\begin{centering}
\includegraphics[viewport=0bp 0bp 1092bp 616bp,clip,width=1\textwidth]{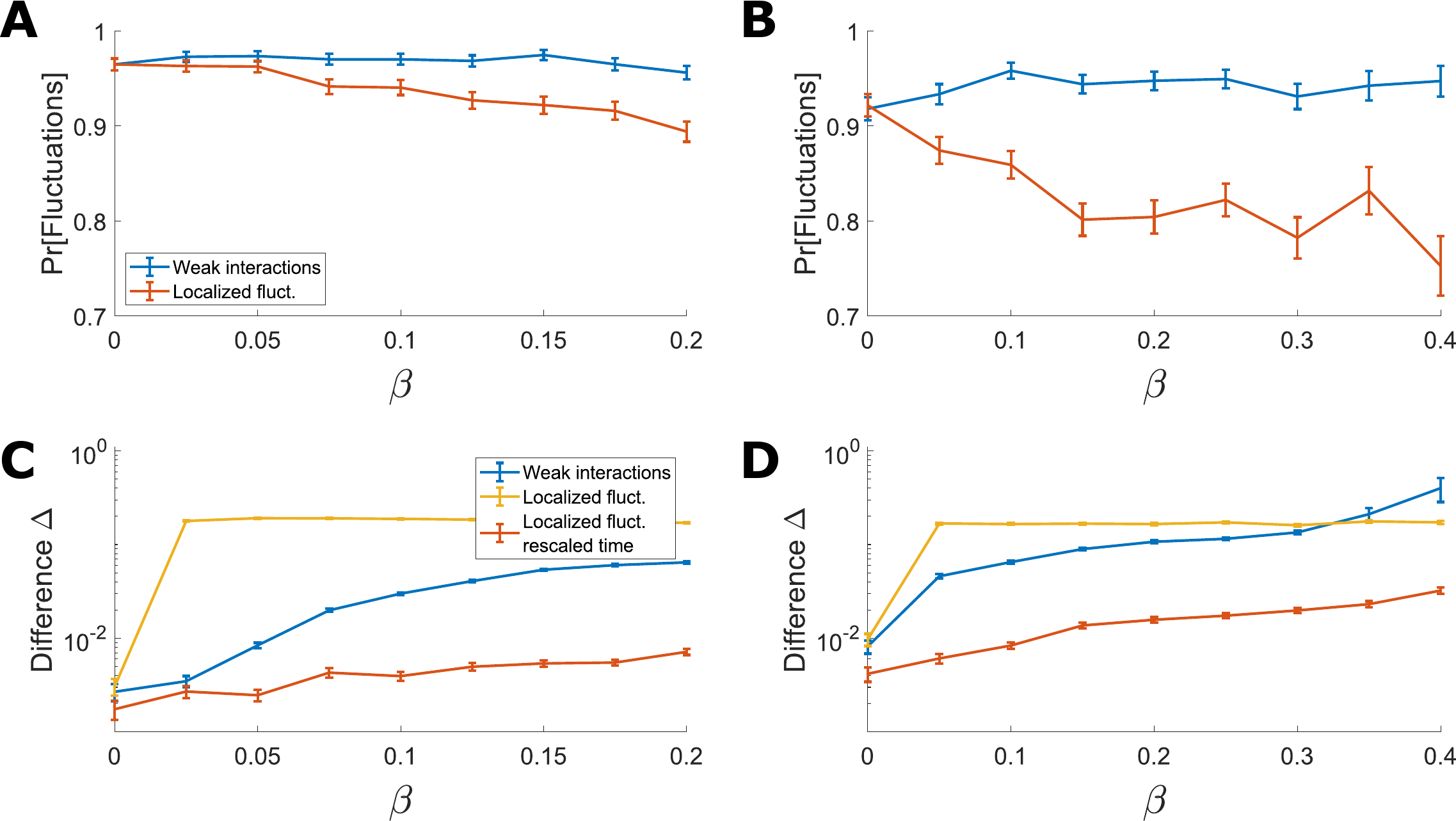}
\par\end{centering}
\caption{\label{fig:beta-dependence}\textbf{The accuracy two types of approximations.}
Shown are measures of the accuracy of the approximate dynamics of
a focal group comprising a directed cycle of length 3 and interaction
strength $\alpha_{\text{focal}}=2.5$, embedded in many different
realizations of the community interactions. The parameters for the
effective dynamics are found using both the weak interactions approximation
and the localized fluctuations approximation at distance $D=0$. Results
are shown for communities with $\alpha=0.2$ in \textbf{(A,C)} and
$\alpha=0.4$ in \textbf{(B,D),} as a function of the reciprocal interaction
strength $\beta\protect\leq\alpha$ \textbf{(A-B) }In cases where
the dynamics of the embedded focal element fluctuate, the probability
of the effective dynamics to reach a fluctuating state for the weak
interactions approximation (blue) and the localized fluctuations approximation
(red). \textbf{(C-D) }The difference between the embedded dynamics
and the effective dynamics of the focal species, taken for the weak
interactions approximation (blue), the localized fluctuations approximation
without time rescaling (yellow) and the localized fluctuations approximation
with time rescaling (red). Further details on parameters are given
in Appendix \ref{sec:Simulation-details}.}
\end{figure*}

\subsubsection{Weak interactions approximation}

Here we will find an approximation for the embedded dynamics that
uses the strength of the interactions in the surrounding community
as an expansion parameter. Thus, this approximation is expected to
work best for weak interactions $\alpha_{ij}$ between the non-focal
species, and between focal and non-focal species. The strength of
the interactions between the focal species is not assumed to be weak.
We also assume that the growth rates $\rho_{k}$ of on-focal species
interacting directly with the focal group not be too slow relative
to the frequency of the focal species dynamics. We look at the effect
of introducing the focal group to the community on a non-focal species
$k$ as $N_{k}(t)=N_{k}^{0}+n_{k}(t)$, where $N_{k}^{0}$ is the
constant value of the abundance of $i$ when the focal species are
removed. As the distance from the focal group grows, $n_{k}$ decreases
by at least one factor of $\alpha_{ij}$ at each step, and the feedback
to the element decreases by another factor of $\alpha_{ji}$. We wish
to approximate the dynamics of the focal species to order $O(\alpha_{ij}^{2})$,
so for species interacting directly with the focal group it is enough
to approximate $n_{k}(t)$ to first order in the $\alpha_{ij}$. For
all other species, further away from the focal group, it is enough
to take $N_{k}\approx N_{k}^{0}$. As detailed in Appendix \ref{sec:Appendix weak interactions approx},
using this approximation for a species $k$ that interacts directly
with a focal species $i$, one finds that $n_{k}(t)\approx-\alpha_{ki}N_{i}(t)+O(\alpha_{ij}^{2})$.
From here, one gets a differential equation for the focal species
dynamics, which is of the from of the Lotka-Volterra equations in
Eq. \ref{eq:LV} with the effective parameters
\begin{align}
K_{i}^{\text{eff}} & =\frac{K_{i}-\sum_{k}\alpha_{ik}N_{k}^{0}}{1-\sum_{k}\alpha_{ik}\alpha_{ki}}\nonumber \\
\rho_{i}^{\text{eff}} & =\rho_{i}\left(1-\sum_{k}\alpha_{ik}\alpha_{ki}\right)\label{eq:weak interactions approximation}\\
\alpha_{ij}^{\text{eff}} & =\frac{\alpha_{ij}}{1-\sum_{k}\alpha_{ik}\alpha_{ki}};i\neq j\nonumber 
\end{align}
where the sums over $k$ run over all non-focal species that are not
extinct in the absence of the focal species.

We show simulation results from systems with constant interaction
strengths $\alpha$ and $\beta$ in Fig. \ref{fig:beta-dependence}.
In cases where the embedded dynamics fluctuate, this approximation
reproduces very well the probability of fluctuations in the range
of parameters that we tested. The measure of dynamical difference,
$\Delta$, is also surprisingly small. (To judge the quality of an
approximation from $\Delta$, it is useful to compare to its value
for unrelated dynamics. As abundances for competitive systems have
$0\leq N_{i}\leq K_{i}$, for $K_{i}=1$ the difference between two
randomly drawn abundances is $\Delta=0.5$.) Furthermore, qualitative
features of the embedded dynamics appear to be well-captured, compare
for example Fig. \ref{fig:setup}(A) and \ref{fig:setup}(D).

The difference becomes larger as $\alpha,\beta$ are increased. Yet,
since the expansion above is exact to order $O(\alpha_{ij}^{2})$
which means $O(\alpha\beta)$ in this case, the approximation is quite
good even for moderate interaction strengths, for example $\Delta<0.1$
for $\alpha=0.4$ and $\beta=0.15$, see Fig. \ref{fig:beta-dependence}(D).

Note that for unidirectional interactions, the effective parameters
given in Eq. \ref{eq:weak interactions approximation} are the same
as the unidirectional approximation given in Eq. \ref{eq: k_eff constant}.
The interaction strengths and growth rates are unchanged, since for
any focal species $i$ and neighboring species $k$, one has $\alpha_{ik}\alpha_{ki}=0$.
The expression for the effective carrying capacities also does not
change: as cycles are long, the abundance of a species $k$ with an
incoming interaction to a focal species $i$ (i.e., with $\alpha_{ik}\neq0$)
are unaffected by the introduction of the focal group, so $N_{k}^{0}=N_{k}$.
The simulations results for the unidirectional case, $\beta=0$, do
not have $\Delta=0$ as expected for $S\to\infty$, since for finite
$S$ short cycles near the focal group generate feedback on the fluctuations.
As we show in Fig. \ref{fig:S dependence} in the appendix, the difference
indeed decays with system size for $\beta=0$, but not for $\beta\neq0$.

The analytical expressions given here for $K_{i}^{\text{eff}},\rho_{i}^{\text{eff}},\alpha_{i}^{\text{eff}}$
involve the parameters $K_{i},\rho_{i}$ and $\alpha_{ij}$ as well
as abundances in the absence of the focal group $N_{k}^{0}$. Directly
using the full expressions to fit a model might therefore be of limited
practical use in situations where they are hard to measure. However,
they are still useful in relating the many of the parameters of the
entire system, and the parameters of the dynamics of the focal group.
More importantly, the existence of these expressions ensures that\emph{
it is possible to fit effective parameters that capture the dynamical
behavior}, at least when the interactions with the rest of the community
are weak enough.

\subsubsection{Localized fluctuations approximation}

\begin{figure*}
\begin{centering}
\includegraphics[width=1\textwidth]{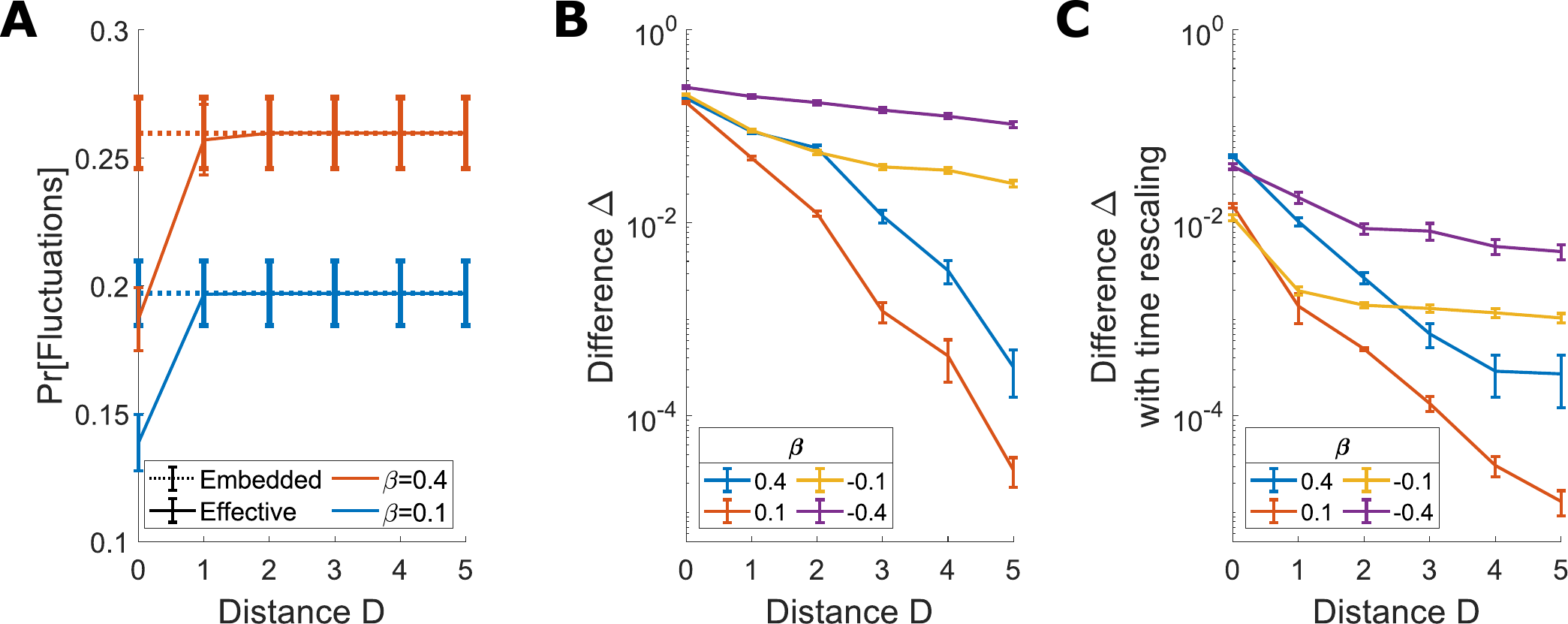}
\par\end{centering}
\caption{\label{fig:statistics}\textbf{The localized fluctuations approximation
improves as the distance from the focal group is increased.} A set
of focal species which make up a unidirectional cycle of length 3
is embedded in a community with bidirectional interactions, with $\alpha=0.7$
and different values of $\beta$. The effective dynamics are approximated
using the localized fluctuations approximation at increasing distances
$D$ from the focal species. Shown are averaged simulation results
for many realization of the interaction matrices. \textbf{(A)} The
probability of the focal species dynamics to fluctuate, when embedded
in the community (dotted) and for the effective few-variable systems
(full). Different colors show different values of $\beta$. As $D$
increases, the probability for the effective dynamics to fluctuate
approaches that of the true, embedded dynamics, and they become equal
for large enough $D$.\textbf{ (B-C)} The difference $\Delta$ between
the embedded dynamics of the focal species and their effective dynamics
as a function of distance $D$. $\Delta$ is calculated over the focal
species only, for cases where the \emph{embedded} dynamics fluctuates.
Results when time is not rescaled are shown in \textbf{(B)}, and for
the same effective dynamics with rescaled time in \textbf{(C).} Further
details on parameters are given in Appendix \ref{sec:Simulation-details}.}
\end{figure*}

Here we introduce a few-species description that includes the focal
group along with its neighbors. We rely on the fact that the fluctuation
amplitudes decay exponentially with the distance from the focal group
in the interaction graph; the fluctuations are therefore localized
around the focal group. Therefore, a good approximation might be to
assume that abundances at distance $D+1$ are constants, then explicitly
model all species up to distance $D$, which we will refer to the
as the focal neighborhood. This amounts to fixing effective carrying
capacities $K^{\mathrm{eff}}$ for the species exactly at distance
$D$, at the edge of the neighborhood. The accuracy of this approximate
description quickly improves, and as much as desired, by increasing
$D$.

The effective carrying capacities $K^{\mathrm{eff}}$ values are obtained
from the abundances $\left\{ N_{i}(t)\right\} $ of the neighborhood
species, \emph{including the focal species}, as follows: at long times,
the embedded dynamics reach a limit cycle. The parameters $K_{i}^{\text{eff}}$
are then taken, similarly to the unidirectional case in Eq. \ref{eq: k_eff constant},
to be 
\begin{equation}
K_{i}^{\text{eff}}=K_{i}-\sum_{j\text{\ensuremath{\notin}neighborhood}}\alpha_{ij}N_{j}(t_{i})\label{eq:k_eff external definition}
\end{equation}
with $t_{i}$ a time in the limit cycle when $N_{i}$ attains a maximum.
As $dN_{i}(t_{i})/dt=0$, and $N_{i}(t_{i})\neq0$, from the LV equations
\ref{eq:LV}, the growth rate must be zero, $g_{i}\left(\vec{N}(t_{i})\right)=0$.
This means that $K_{i}^{\text{eff}}$ can also be expressed using
only the abundances in the focal neighborhood, as
\begin{equation}
K_{i}^{\text{eff}}=\sum_{j\in\text{neighborhood}}\alpha_{ij}N_{j}(t_{i})\,.\label{eq: k_eff internal definition}
\end{equation}
As the species at the boundary of the neighborhood, at distance $D$,
are the only explicitly modeled species that interact with any species
outside of the chosen neighborhood, all species at smaller distances
will have $K_{i}^{\text{eff}}=K_{i}$.

Fig. \ref{fig:statistics}(B) shows that the difference between the
effective and embedded dynamics, $\Delta$, decays exponentially with
distance $D$ for several values of $\alpha$ and $\beta$. (Note
that $\Delta$ is calculated only over the focal species and not the
entire neighborhood.) The probability of the effective dynamics to
reach a fluctuating state also converges quickly to the that of the
embedded dynamics as $D$ is increased, see Fig. \ref{fig:statistics}(A).
See also Fig. \ref{fig:appendix statistics other types of interactions}
in the appendix for statistics on other types of focal groups. Recall
that if the dynamics of the embedded focal species is at a fixed-point,
the effective description is exact and will also reach a fixed point;
therefore, if the probability of fluctuations is identical for the
effective and embedded dynamics, the cases where the effective dynamics
fluctuate are exactly the ones where the embedded dynamics do so.

In practice, in many cases a good approximation of the embedded dynamics
can be achieved even for small $D$. This is true even for $D=0$,
where only the focal species are explicitly modeled. Interestingly,
the effective description at low $D$ can be improved by rescaling
time, as can be seen by comparing Fig. \ref{fig:statistics}(B) and
Fig. \ref{fig:statistics}(C), as well as in Fig. \ref{fig:beta-dependence}(C-D).
In appendix \ref{sec:appendix time rescaling}, we show that on a
simple system the time rescaling in necessary in order to fix both
the coefficients of $N_{i}$ and $N_{i}^{2}$ in the Lotka-Volterra
equation. In many cases, after the rescaling the effective few-species
dynamics become almost identical to the embedded dynamics already
for $D=0$, see examples in Fig. \ref{fig:examples}. Without adding
the time rescaling, as the frequency of the effective and embedded
dynamics are different, at long times the dynamics become completely
asynchronized and the difference $\Delta$ between them is the same
as if they were random. This can be seen in Fig. \ref{fig:beta-dependence}(B,D),
showing that without time-rescaling $\Delta$ becomes large and independent
of $\beta$ for any $\beta>0$. This is due to the change in frequency
rather than other differences, as $\Delta$ becomes greatly improved
and $\beta$-dependent when the frequency is adjusted.

\section{Discussion}

We ask when and how it is possible to approximate the dynamics of
a small set of focal species, forming part of a large community, using
few-variable dynamical models. Within the Lotka-Volterra framework,
we study small subsets of focal species whose abundances fluctuate
in isolation, embedded in large, sparse ecological networks. We find
that in dynamical phases where significant fluctuations in the embedding
community encompass at most a finite number of species, it is often
possible to approximate the dynamics of the embedded element using
a model with only several species. However, in phases where a finite
fraction of species abundances fluctuate, the dynamics are high-dimensional
and cannot be modeled by a small number of variables.

If interactions are unidirectional, the abundances of species affecting
the focal species are fixed in the limit of many species, and so the
community affects the focal species only through a constant change
to the carrying capacities. The dynamics can therefore always be modeled
exactly with the Lotka-Volterra equations, considering the only focal
species and using adjusted parameters. When interactions are bidirectional,
an exact description is not always possible due to feedbacks. Here,
we offer two methods of finding a few species description. In one
approach, we assume that all interactions in the community are very
weak, and the introduction of the focal species affects it only as
a small perturbation. With this assumption, we find effective values
up to second order in the interaction strengths, to the parameters:
interaction strengths, growth rates and carrying capacities. In the
other approach, we assume that fluctuations decay quickly with distance
from the focal group, so that at some distance $D+1$ the abundances
can be approximated as fixed. Dynamics are then approximated by explicitly
modeling the finite number of species that are at up to distance $D$
from the focal species, using changes to the carrying capacities similar
to the ones done in the unidirectional case. This approximation can
be improved as much as desired by increasing $D$; however, by adding
a time rescaling, we find that taking $D=0$ and explicitly modeling
the focal species alone often yields a very good approximation of
the embedded dynamics.

The two different approximations are relevant to two different scenarios.
In the first scenario, one observes a small number of species, and
tries to fit these observations to a few-species model. This case
is relevant to the weak interactions approximation, where we find
approximated values for all system parameters. The analytical expressions
giving the effective parameters may be useful if the original parameters
are known. Even if they are not, the very existence of such effective
parameters is important, as these values could be the ones found by
fitting the element dynamics to Lotka-Volterra dynamics. In the second
scenario, one observes in an experiment a small set of species whose
abundances fluctuate in isolation, and measures their interaction
parameters. One then observes these same species (and them alone)
in nature, where there are interactions with many other species, and
sees how the parameters change in the presence of these species. This
is relevant to the localized fluctuations approximation: here we begin
with all true parameters for the element as would be measured in an
isolated experiment. Then, in order to get an effective description
of the embedded dynamics, one needs only to use measurements of the
element species abundances at a few specific times to get effective
carrying capacities.

\appendix
\onecolumngrid

\section{\label{sec:appendix time rescaling}Time rescaling}

Here we will demonstrate, using a simple example, the reason for the
different time scales between the embedded and effective dynamics
when using the localized fluctuations approximations. We will consider
a simple system where each focal species interacts weakly with only
one non-focal species, and with a symmetry between all species. We
will then show how the localized fluctuations approximation can be
related by a time rescaling to the weak interactions approximation,
which approximated the embedded dynamics very well for small $\alpha$.

\begin{figure}[b]
\begin{centering}
\includegraphics[width=1\textwidth]{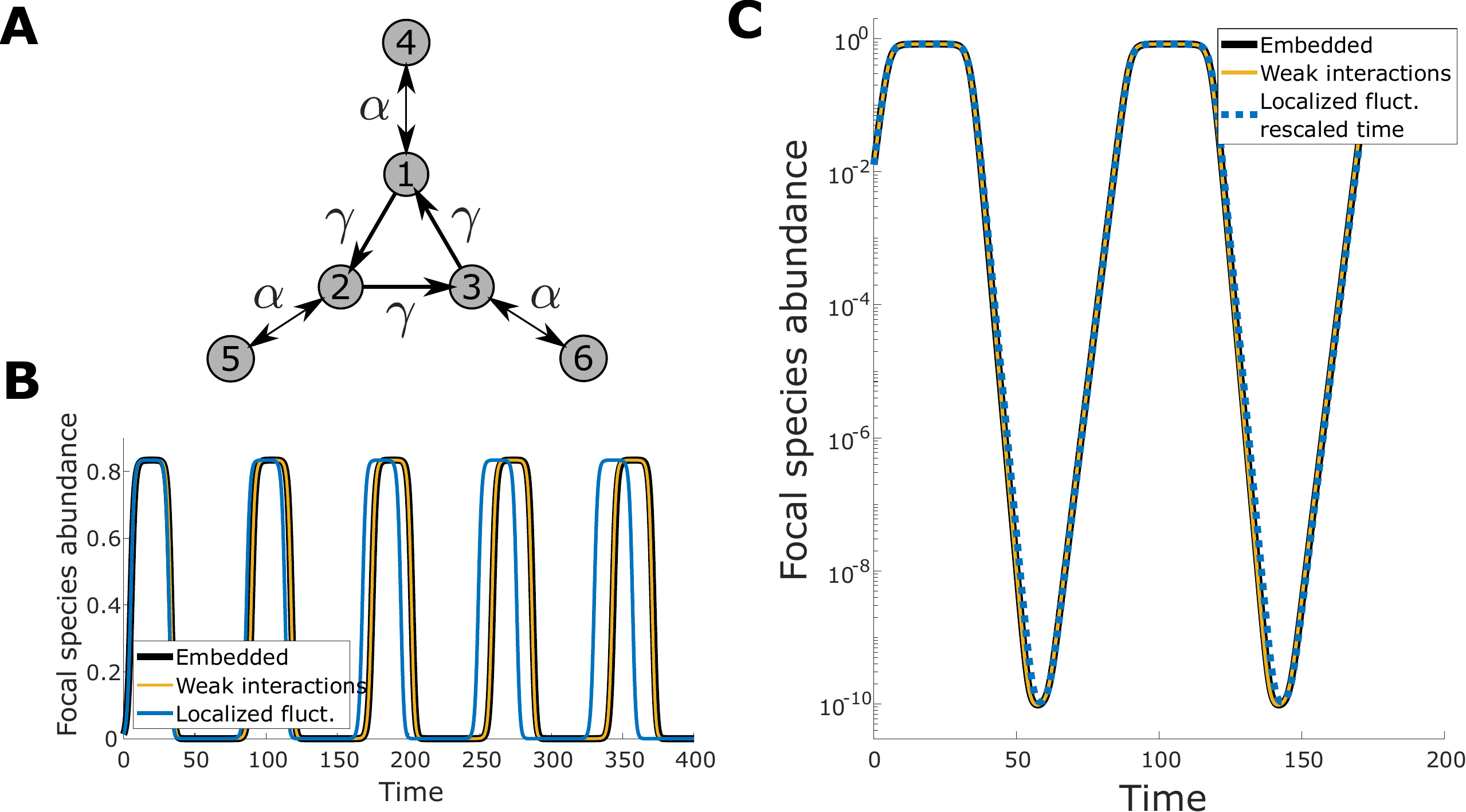}
\par\end{centering}
\caption{\label{fig:cycle with appendages}\textbf{The effect of time rescaling
in a simple case.} \textbf{(A)} The focal species are arranged along
a unidirectional cycle, with interaction strength $\gamma$. Each
one interacts symmetrically with a single non-focal species with strength
$\alpha$. \textbf{(B-C)} Embedded and effective dynamics found for
$\alpha=0.2$, $\gamma=2.3$. \textbf{(B)} Comparison between the
embedded dynamics of the abundance $N_{i}$ of one of the focal species
(thick black line), and the effective dynamics using the weak interactions
approximation (yellow) and the localized fluctuations approximation
without time rescaling (blue). \textbf{(C)} Comparison between the
embedded dynamics of the abundance $N_{i}$, shown in log scale, of
one of the focal species (thick black line), and the effective dynamics
using the weak interactions approximation (yellow) and the localized
fluctuations approximation with time rescaling (blue, dashed).}
\end{figure}

We will consider the case illustrated in Fig. \ref{fig:cycle with appendages}A.
The focal group make up a directed, unidirectional cycle of odd length
$n$, where each species $i$ affects the next species $i+1$ with
an interaction strength $\gamma>2.5$, so that in isolation the dynamics
would fluctuate by switching between $N_{i}=0,1$. Each focal species
$i$ also interacts with one non-focal neighbor $i+n$, with a bidirectional
interaction of strength $\alpha$ (so $\alpha_{i,i+n}=\alpha_{i+n,i}=\alpha$).
There are no other interactions in the system. For all species, $K_{j}=1,\,r_{j}=1$.
Thus, the LV equations for species $i$ and $i+n$ would be 
\begin{equation}
\frac{dN_{i}}{dt}=N_{i}\left(1-N_{i}-\gamma N_{i-1}-\alpha N_{i+n}\right)+\lambda\label{eq:cycle species LV}
\end{equation}
\begin{equation}
\frac{dN_{i+n}}{dt}=N_{i+n}\left(1-N_{i+n}-\alpha N_{i}\right)+\lambda\label{eq:appendage species LV}
\end{equation}

Noting that the abundances of the non-focal species would be $N_{i+n}^{0}=1$
in the absence of the focal species (as they interact with no other
species), using Eq. \ref{eq:weak interactions approximation} for
the parameters of the weak interactions approximation, one gets $K_{i}^{\text{eff}}=1/\left(1+\alpha\right),\,\rho_{i}^{\text{eff}}=\rho_{i}\left(1-\alpha^{2}\right),\,\gamma^{\text{eff}}=\gamma/\left(1-\alpha^{2}\right)$.
The effective LV equations for the system found by using the weak
interactions approximation, $\left\{ N_{i}^{\text{WI}}\right\} $,
are therefore 
\begin{equation}
dN_{i}^{\text{WI}}/dt=N_{i}^{\text{WI}}\left[\left(1-\alpha\right)-\left(1-\alpha^{2}\right)N_{i}^{\text{WI}}-\gamma N_{i-1}^{\text{WI}}\right]+\lambda\label{eq: cycle weak-interactions approx}
\end{equation}
See comparison of these effective dynamics for 3 focal species \ref{eq: cycle weak-interactions approx}
to the embedded dynamics from the full 6-species community in Fig.
\ref{fig:cycle with appendages}(B-C).

We now wish to compare these result to the effective dynamics found
using the localized fluctuations approximation. For each focal species
$i$, we take $K_{i}^{\text{eff}}=K_{i}-\alpha N_{i+n}(t_{i})$, where
$t_{i}$ is the time when $N_{i}$ is at a maximum. At this time,
its growth rate must have $g_{i}(\vec{N})=0$. The abundance of the
incoming focal species along the cycle, $i-1$, can be approximated
as $N_{i-1}\approx0$. We can also approximate that at this time,
the non-focal species $i+n$ is at a minimum, so $dN_{i+n}/dt\approx0$,
which means that $g_{i+n}(\vec{N})\approx0$ (as $\alpha$ is small,
species $i$ cannot drive species $i+n$ to extinction, meaning that
$N_{i+n}>0$). In total, this means that 
\begin{align*}
g_{i}\left(\vec{N}(t_{i})\right) & \approx1-N_{i}(t_{i})-\alpha N_{i+n}(t_{i})\approx0\\
g_{i+n}\left(\vec{N}(t_{i})\right) & =1-N_{i+n}(t_{i})-\alpha N_{i}(t_{i})\approx0
\end{align*}
Using both equations, we get that $N_{i+n}(t_{i})\approx1/\left(1+\alpha\right)$,
and so $K_{i}^{\text{eff}}=1/\left(1+\alpha\right)$. We now also
add a new timescale $\tau=\left(1-\alpha^{2}\right)t$, so that the
LV equations for the effective system found using the localized fluctuations
approximation with the time rescaling $\left\{ N_{i}^{\text{LF}}\right\} $
behaves as 
\[
\frac{dN_{i}^{\text{\text{LF}}}}{d\tau}=N_{i}^{\text{\text{LF}}}\left(K_{i}^{\text{eff}}-N_{i}^{\text{\text{LF}}}-\gamma N_{i-1}^{\text{\text{LF}}}\right)+\lambda
\]
which finally gives the equation
\begin{equation}
\frac{dN_{i}^{\text{\text{LF}}}}{dt}=N_{i}^{\text{\text{LF}}}\left(\left(1-\alpha\right)-\left(1-\alpha^{2}\right)N_{i}^{\text{\text{LF}}}-\left(1-\alpha^{2}\right)\gamma N_{i-1}^{\text{\text{LF}}}\right)+\left(1-\alpha^{2}\right)\lambda\label{eq:cycle weak-fluctuations approx}
\end{equation}
Comparing equations \ref{eq: cycle weak-interactions approx} and
\ref{eq:cycle weak-fluctuations approx}, we can see that they differ
only in taking $\gamma\rightarrow\left(1-\alpha^{2}\right)\gamma$
and $\lambda\rightarrow\left(1-\alpha^{2}\right)\lambda$. Fig. \ref{fig:cycle with appendages}(C)
shows that he dynamics are almost identical, both to each other and
to the embedded dynamics. The correction to $\gamma$ affects the
dynamics only by changing slightly the rate of approach of $N_{i}$
to it lowest $O(\lambda)$ values, an effect that can only be discerned
when $N_{i}(t)$ is many orders of magnitude smaller than $1$ and
is effectively extinct. As $\lambda$ is very small, the correction
to migration also has little effect on the dynamics.

On the other hand, if we do not use time-rescaling, we get that in
the LV equation neither the coefficients of $N_{i}$ nor $N_{i}^{2}$
fit the weak-interactions equations. Indeed, as can be seen in Fig.
\ref{fig:cycle with appendages}(B), this causes the effective dynamics
to fluctuate much faster that the embedded dynamics.

\section{\label{sec:Appendix weak interactions approx}Weak interactions approximation}

Here we will derive the expressions for the effective parameters in
the weak interactions approximation. We will assume that all interactions
in the embedding community are $\ll1$, although the interactions
for the focal species can be large. We will also assume that the growth
rates $\rho_{i}$ for the non-focal species interacting directly with
the focal group are fast relative to the embedded dynamics.

\begin{figure}
\begin{centering}
\includegraphics[width=0.65\columnwidth]{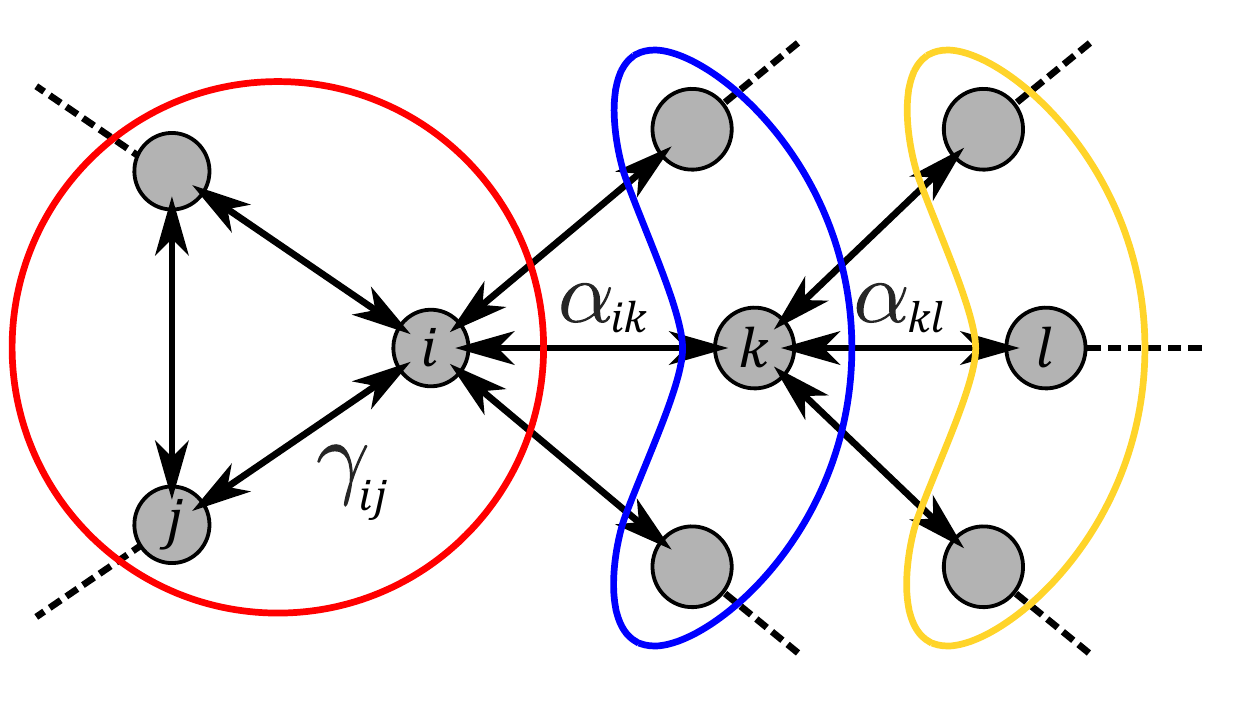}
\par\end{centering}
\caption{\label{fig:approximation calculation explanation}\textbf{The focal
species and their neighborhood, as used in the calculation of the
effective values of the weak interactions approximation.} The focal
species (in this example, making up a cycle) are denoted as $i$ and
$j$, and interact with $\gamma_{ij}$. They makes up the zeroth layer
and are circled in red. The first layer species, the ones which interact
directly with the focal species, are circled in blue and are labeled
using $k$. They interact with the focal species with $\alpha_{ik}$.
The second layer species, which interact with the first layer, are
circled in yellow and are labeled using $l$. They interact with the
first layer with $\alpha_{kl}$.}
\end{figure}

For a given set of focal species, we will consider three ``layers''
of species at increasing distances from the focal group, see Fig.
\ref{fig:approximation calculation explanation}. In the 0th layer,
the focal group, species will be labeled as $i$ or $j$, and the
interactions between them as $\gamma_{ij}$. In the 1st layer, species
directly interacting with the focal group, species will be labeled
$k$, and the 2nd layer, species interacting with the first layer,
species will be labeled $l$. All interactions that are not between
two 0th layer species have values drawn from the wider community distribution,
so they will have $\alpha_{ik},\alpha_{kl}\ll1$. As interactions
are sparse, loops are rare and we can assume that there are no species
that belong to several layers at once. We will also assume that none
of the species are extinct in the absence of the focal species: such
species will remain extinct when the focal group is introduced (up
to the order $O(\alpha_{ik}^{2})$ in which we are interested) and
so will have no effect on the focal species. Therefore we can effectively
remove them from the system.

We will consider the introduction of the focal group as a perturbation
in the rest of the system. Defining $N_{k}^{0},N_{l}^{0}$ as the
(constant) values of the abundances of first and second layer species
when the focal species are removed from the system, we will find the
small corrections to the abundances $N_{k}(t)-N_{k}^{0}$ as a power
series in the community interaction strengths.

For a focal species $i$, the equation for $dN_{i}/dt$ is given by
\begin{equation}
dN_{i}/dt=\rho_{i}N_{i}\left(K_{i}-N_{i}-\sum_{j\in\text{0th layer}}\gamma_{ij}N_{j}-\sum_{k\in\text{1st layer}}\alpha_{ik}N_{k}\right)+\lambda_{i}\label{eq:zeroth layer}
\end{equation}
From here, the correction to the abundance $N_{k}(t)-N_{k}^{0}$ of
the first layer species need be approximated only up to first order
in the interaction strengths, as they only appear in the equation
when multiplied by $\alpha_{ik}$.

Consider a first layer species $k$ interacting with the element species
$i$. In addition, it can also interact with second layer species,
$l$. Its dynamics are therefore given by 
\begin{equation}
dN_{k}/dt=\rho_{k}N_{k}\left(K_{k}-N_{k}-\alpha_{ki}N_{i}-\sum_{l\in\text{2nd layer}}\alpha_{kl}N_{l}\right)+\lambda_{k}\label{eq:first layer}
\end{equation}
As $N_{l}$ appears in the equation already multiplied by $\alpha_{kl}$,
it is enough to approximate it to zeroth order, and take $N_{l}=N_{l}^{0}+O(\alpha_{ij})$.
Now, denoting 
\begin{align}
N_{k}(t) & =N_{k}^{0}-\alpha_{ki}n_{k}(t)+O\left(\alpha_{ij}^{2}\right)\,,
\end{align}
and plugging into Eq. \ref{eq:first layer}, one gets that
\begin{align}
dn_{k}/dt & =\rho_{k}N_{k}^{0}\left(-n_{k}+N_{i}(t)\right)+\lambda_{k}+O(\alpha_{ik}^{2})
\end{align}
where $N_{k}^{0},N_{l}^{0}$ were related through the fixed point
equation for $N_{k}^{0}$ in the absence of the element: $N_{k}^{0}=K_{k}-\sum_{l}\alpha_{kl}N_{l}^{0}$.
If $N_{i}(t)$ is slow relative to $\rho_{k}$, one can take $\dot{n}_{k}\approx0$,
so $n_{k}=N_{i}(t)$ and $N_{k}(t)=N_{k}^{0}-\alpha_{ki}N_{i}(t)$.

Plugging $N_{k}\left(t\right)$ back into the zeroth layer equation
\ref{eq:zeroth layer}, the equation becomes
\begin{align}
dN_{i}/dt & =\rho_{i}N_{i}\left[\left(K_{i}-\sum_{k}\alpha_{ik}N_{k}^{0}\right)-\left(1-\sum_{k}\alpha_{ik}\alpha_{ki}\right)N_{i}-\sum_{j}\gamma_{ij}N_{j}\right]+\lambda_{i}+O\left(\alpha_{ik}^{2},\alpha_{ik}\lambda_{k}\right)
\end{align}
which is the same as the LV equations when taking 
\begin{align}
K_{i}^{\text{eff}} & =\frac{K_{i}-\sum_{k}\alpha_{ik}N_{k}^{0}}{1-\sum_{k}\alpha_{ik}\alpha_{ki}}\nonumber \\
\rho_{i}^{\text{eff}} & =\rho_{i}\left(1-\sum_{k}\alpha_{ik}\alpha_{ki}\right)\\
\gamma_{ij}^{\text{eff}} & =\frac{\gamma_{ij}}{1-\sum_{k}\alpha_{ik}\alpha_{ki}};i\neq j\nonumber 
\end{align}

\begin{figure}
\begin{centering}
\includegraphics[width=1\textwidth]{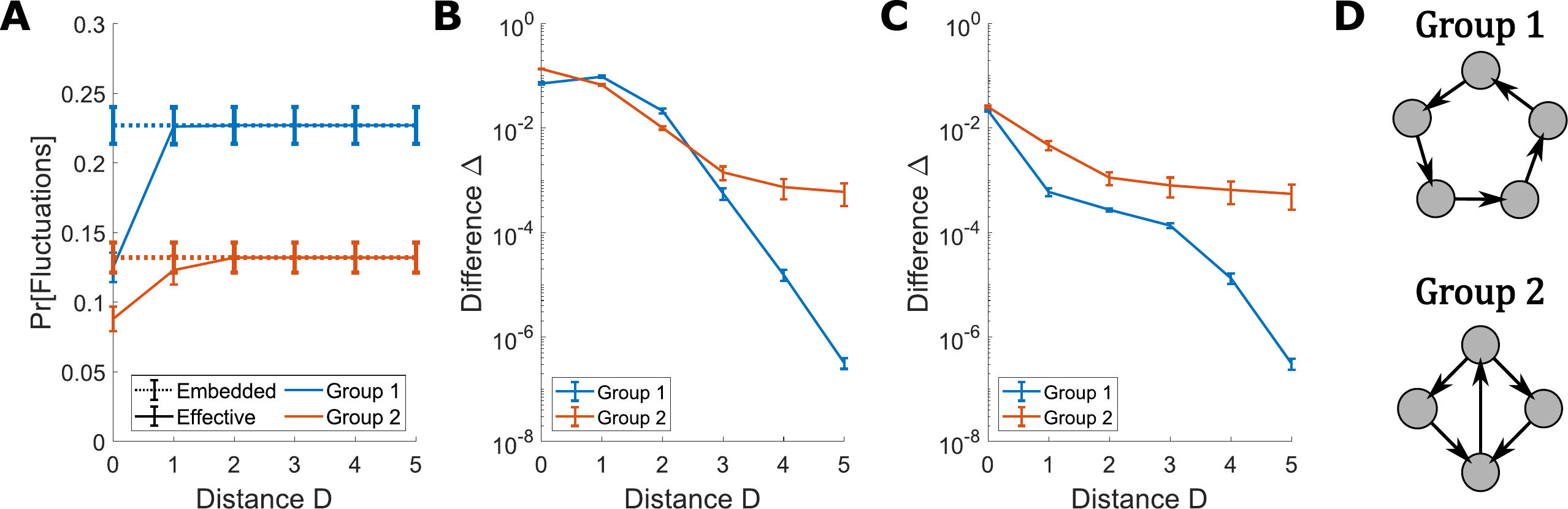}
\par\end{centering}
\caption{\label{fig:appendix statistics other types of interactions}\textbf{The
localized fluctuations approximation improves as the distance grows
for more focal groups.} Two different focal groups, shown in \textbf{(D)},
are embedded in communities with $\alpha=0.4,\,\beta=0.1$. Effective
dynamics are found using the localized fluctuations approximation
at increasing distances $D$ from the focal species. Shown are averaged
simulation results for many realization of the interaction matrices.
\textbf{(A)} The probability of the elements to fluctuate, when embedded
in the community (dotted) and for the effective few-variable systems
(full). Different colors represent the two different focal groups.\textbf{
(B-C)} The difference $\Delta$ between the embedded and effective
dynamics as a function of distance $D$. $\Delta$ is calculated over
the focal species only, for cases where the \emph{embedded} dynamics
fluctuates. Results where time is not rescaled are shown in \textbf{(B)},
and for the same dynamics with rescaled time in \textbf{(C)}.}
\end{figure}

\begin{figure}
\begin{centering}
\includegraphics[width=0.8\columnwidth]{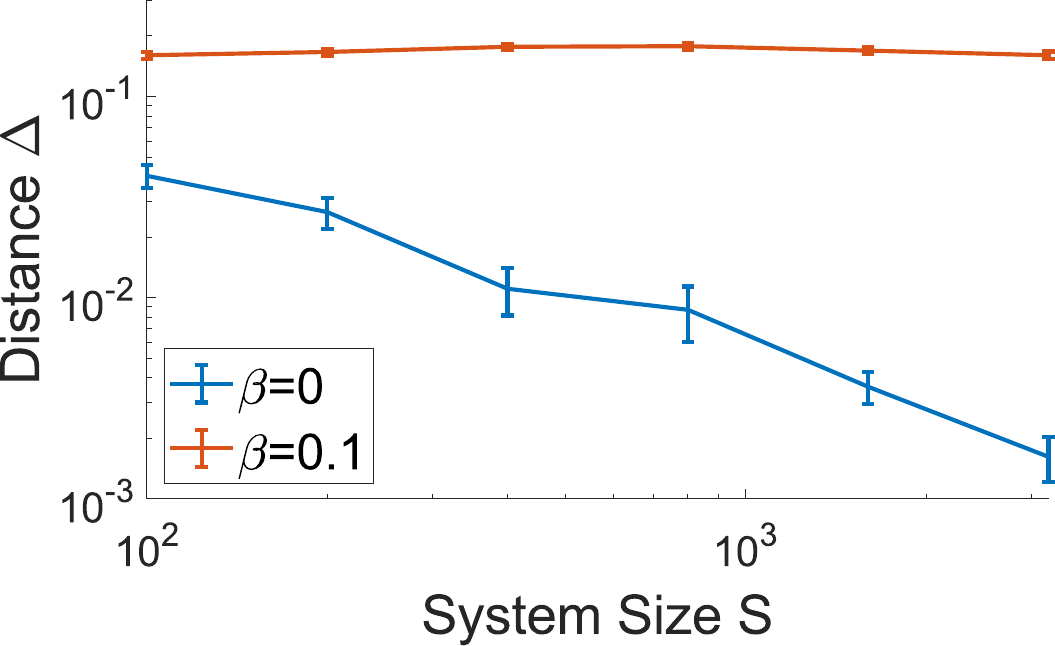}
\par\end{centering}
\caption{\label{fig:S dependence}\textbf{The localized fluctuations approximation
improves with system size for unidirectional interactions.} A focal
group making up a length-3 unidirectional cycle with interaction strength
$\alpha_{\text{focal}}=2.5$ is embedded in communities with $C=3$
and $\alpha=0.4$. For unidirectional interactions $\beta=0$, (blue)
and for bidirectional interactions $\beta=0.1$ (red). Shown is the
distance $\Delta$ between the embedded and effective dynamics, using
the localized fluctuations approximations with no time rescaling,
as a function of the embedding system size $S$. $\Delta$ is calculated
in cases where the embedded dynamics fluctuate. In the unidirectional
case, the distance decreases with system size, and in the limit $S\rightarrow\infty$
the approximation becomes exact. For the bidirectional case, the approximation
does not improve with $S$.}
\end{figure}

\section{\label{sec:Simulation-details}Simulation details}

Here we have the details of simulations, and exact parameters for
the graphs appearing in the main text. For results involving statistics
over many realizations, such as Figs. \ref{fig:beta-dependence},\ref{fig:statistics},
at each realization we independently sample the interaction matrix
$\alpha_{ij}$ of the embedding community. We then connect each focal
species to any non-focal species with probability $C/2S$, again sampled
independently. For each system, we solve the LV equations using an
ODE45 solver, with $K_{i}=1,\rho_{i}=1$ and $\lambda_{i}=10^{-10}$.
The initial conditions are randomly sampled from a uniform distribution
over $\left[0,1\right]$ for each $N_{i}$. The dynamics are first
run for $t_{\text{initial}}=1000/\rho_{i}$, when we assume that they
have relaxed to either a fixed point or a limit cycle, and then for
a further $t_{\text{run}}=10000/\rho_{i}$. We then do the same in
the effective system.

In order to calculate the difference $\Delta$, we need to compare
the embedded dynamics $N_{i}^{\text{embedded}}\left(t\right)$ and
the effective dynamics $N_{i}^{\mathrm{eff}}\left(t\right)$. As both
are given time to relaxation, the effective dynamics, even if exactly
capturing the embedded dynamics, can be shifted relative to them.
In order to find this shift, we assume that $N_{i}$ can be considered
linear between any sampled timepoints, then find the shift that yields
the minimal $\Delta$.

In Figs. \ref{fig:beta-dependence},\ref{fig:statistics},\ref{fig:examples}
in the main text and \ref{fig:appendix statistics other types of interactions}
in the appendices, we use systems with $S=2000,C=3$. In Fig. \ref{fig:examples}
of the main text, each set of focal species is embedded in a community
with $S=2000$, $C=3$, $\alpha=0.4$, $\beta=0.1$. In order to see
that the examples are not unusual cases, we compare the distance in
each example to the distances found when running 1000 realizations,
and taking the cases where the embedded dynamics fluctuate. In the
first example, the focal elements make up a cycle of length three
with interaction strengths $\alpha_{\text{focal}}=2.5$. It has $\Delta\approx0.0043$,
larger than the distance in 69\% of realizations. In the second example,
the focal elements make up a cycle of length five with $\alpha_{\text{focal}}=1.3$.
It has $\Delta\approx0.0043$, larger than the distance in 49\% of
realizations. In the third example, the focal elements make up two
cycles of length three that share a side, with $\alpha_{\text{focal}}=1.8$.
It has $\Delta\approx0.005$, larger than the distance in 36\% of
realizations.

\bibliographystyle{unsrt}
\bibliography{local_description}

\end{document}